\documentclass{PoS}

\title{$\mathbf{U_A(1)}$ breaking at finite temperature from the Dirac spectrum with the dynamical HISQ action}

\ShortTitle{$U_A(1)$ breaking at finite temperature from the Dirac spectrum with the dynamical HISQ action}

\author{
\speaker{H.~Ohno},$^a$ U.M.~Heller,$^b$ F.~Karsch,$^{a,c}$ and S.~Mukherjee$^c$ \\
\llap{$^a$}Fakult\"{a}t f\"{u}r Physik, Universit\"{a}t Bielefeld, \\
D-33615 Bielefeld, Germany \\
\llap{$^b$}American Physical Society, \\
One Research Road, Ridge, NY 11961, USA \\
\llap{$^c$}Physics Department, Brookhaven National Laboratory, \\
Upton, NY 11973, USA \\
E-mail: \email{hono@physik.uni-bielefeld.de}
}

\abstract{
We investigate $U_A(1)$ breaking above $T_c$ in terms of the Dirac spectrum on configurations
with (2+1)-flavors, using the HISQ action. The strange quark mass is at its physical value.
We use several light quark masses corresponding to the Goldstone pion masses in the range of about 115 -- 230 MeV
on lattices of size 32$^3 \times$8 and 48$^3 \times$8.  We calculate the 100 lowest-lying Dirac
eigenvalues at temperatures below and above $T_c$.
We investigate the volume dependence of the Dirac eigenvalue density to determine whether there is
a gap around zero, which can appear if $U_A(1)$ symmetry is restored in the chiral symmetric phase.
We also investigate the quark mass dependence of the Dirac eigenvalue density at zero and check
whether there is a linear behavior that would signal the $U_A(1)$ breaking above $T_c$.
}

\FullConference{
The 30 International Symposium on Lattice Field Theory - Lattice 2012, \\
June 24-29, 2012 \\
Cairns, Australia
}

\begin{document}

\section{Introduction}
Quantum chromodynamics (QCD) with $N_f$ flavors is invariant under a global flavor symmetry of $SU_L(N_f) \times SU_R(N_f) \times U_V(1) \times U_A(1)$ in the
limit of vanishing quark mass. It is also known that the $SU_L(N_f) \times SU_R(N_f)$ chiral symmetry is spontaneously broken in the vacuum and
the $U_A(1)$ symmetry is explicitly broken due to the axial anomaly. However the chiral symmetry is expected to be restored at high temperature.
On the other hand, although the anomaly effect is present at any finite temperature, it is suppressed at high
temperature and accordingly the $U_A(1)$ symmetry is effectively restored.
Here an important question is when and how such an effective $U_A(1)$ symmetry restoration occurs since it can influence the order of the chiral
phase transition. In the case of $N_f=2$ it has been suggested \cite{Pisarski-Wilczek} that if the $U_A(1)$
symmetry remains broken at the critical temperature $T_c$ of the chiral phase transition, the phase transition is of second order
belonging to a universality class of a 3-dimensional $O(4)$ spin model. On the other hand, if both the chiral and the $U_A(1)$ symmetries are restored
at the same time, the phase transition should be of first order.

To study the QCD phase structure and properties of strongly interacting matter, where nonpertabative effects play an important role,
lattice QCD simulations are a suitable way.
Some lattice QCD studies with (2+1)-flavor staggered type quarks \cite{Ejiri:2009ac,Kaczmarek:2011zz} have indicated at least $O(N)$ scaling
in the analysis of the magnetic equation of state, although only the $O(2)$ symmetry is preserved.
Similarly in the case of 2-flavor $O(a)$ improved Wilson quarks \cite{Iwasaki:1996ya,Ali Khan:2000iz} $O(4)$ scaling has been reported.
This suggests the second order phase transition scenario in the case of $N_f=2$ massless quarks.
In our previous report \cite{lattice2011} we investigated the temperature dependence of the Dirac eigenvalue distribution and discussed a scaling behavior
in the (2+1)-flavor case with the highly improved staggered quark (HISQ) action \cite{HISQ}. There we found that the eigenvalue density around zero,
which is related to the chiral order parameter, has a crossover behavior and a power law dependence similar to what is expected from the critical exponent of
the $O(2)$ or $O(4)$ universality classes. Moreover, we found no clear evidence for a gap around zero in the Dirac eigenvalue density at high temperature,
which supports $U_A(1)$ breaking above $T_c$.
Recently the restoration of the $U_A(1)$ symmetry in terms of the Dirac eigenvalue distribution has been discussed also with domain wall \cite{Domain_Wall}
and overlap \cite{Overlap} fermions, which have exact chiral symmetry on the lattice.
With domain wall fermions there remain small eigenvalues even at high temperature. Whereas with overlap fermions a gap around zero seems to appear
above a certain temperature. However, the lattice sizes in these studies are not large enough for definite conclusions and more detailed investigations are needed.

In this report we continue our study on the Dirac eigenvalue distribution at finite temperature with the same setup as in \cite{lattice2011} and give more details about
its volume and quark mass dependences and the effective $U_A(1)$ restoration. In the next section we discuss details of our study.
Then we show some numerical results in Sec. \ref{sec:results} and finally give conclusions in Sec. \ref{sec:conclusions}.

\section{Effective $\mathbf{U_A(1)}$ restoration and the Dirac eigenvalue density}

To discuss the effective restoration of the $U_A(1)$ symmetry, we consider the chiral condensate $\langle \bar{\psi}\psi \rangle$
and several susceptibilities $\chi$ defined in the 4-dimensional Euclidean space-time as follows:
\begin{equation}
\langle \bar{\psi}\psi \rangle \equiv \frac{T}{V}\frac{\partial \ln Z}{\partial m} = \frac{T}{V} \langle \mathrm{Tr}M^{-1} \rangle,
\end{equation}
\begin{equation}
\chi_{\delta} = \chi_{\mathrm{con}} \equiv -\frac{T}{V} \langle \mathrm{Tr}(M^{-1}M^{-1}) \rangle,
\end{equation}
\begin{equation}
\chi_{\mathrm{disc}} \equiv \frac{T}{V} \left [\langle (\mathrm{Tr}M^{-1})^2 \rangle - \langle \mathrm{Tr}(M^{-1}) \rangle^2 \right ],
\end{equation}
\begin{equation}
\chi_{\sigma} = \frac{T}{V}\frac{\partial^2 \ln Z}{\partial m^2} = \frac{\partial \langle \bar{\psi}\psi \rangle}{\partial m} = \chi_{\mathrm{con}} + \chi_{\mathrm{disc}},
\end{equation}
\begin{equation}
\chi_{\pi} \equiv \frac{T}{V} \langle \mathrm{Tr}(M^{-1}\gamma_5 M^{-1}\gamma_5) \rangle,
\end{equation}
where $Z$ is the QCD partition function and $V$, $T$, $m$ and $M$ are volume, temperature, quark mass and the fermion matrix, respectively.
If the chiral symmetry is unbroken, $\langle \bar{\psi}\psi \rangle = 0$ and $\chi_\pi = \chi_\sigma$, and
accordingly $\chi_\pi - \chi_\delta = \chi_{\mathrm{disc}}$. In addition, if the $U_A(1)$ symmetry is also unbroken,
$\chi_\pi = \chi_\delta$ and $\chi_\pi - \chi_\delta = \chi_{\mathrm{disc}} = 0$.
Here $\langle \bar{\psi}\psi \rangle$ and $\chi_\pi - \chi_\delta$ can be given in terms of the Dirac eigenvalue density
$\rho(\lambda,m) \equiv (T/V) \langle \sum_k \delta(\lambda_k(m) - \lambda) \rangle$ as follows:
\begin{equation}\label{pbp}
\langle \bar{\psi}\psi \rangle = \int^{\infty}_0 \frac{2m \rho(\lambda,m)}{\lambda^2 + m^2} d\lambda,
\end{equation}
\begin{equation}\label{ua1}
\chi_\pi - \chi_\delta = \int^{\infty}_0 \frac{4m^2 \rho(\lambda,m)}{(\lambda^2 + m^2)^2} d\lambda,
\end{equation}
where we take $V \rightarrow \infty$. Then, if $m \rightarrow 0$ is also taken, the Banks-Casher relation \cite{Banks-Casher}
$\langle \bar{\psi}\psi \rangle = \pi \rho(0,0)$ follows from (\ref{pbp}), which means $\rho(0,0)=0$ in the chiral symmetric phase.
Thus eigenvalues close to $\lambda=0$ should be suppressed around $T_c$ even if the quark mass is finite.
On the other hand, (\ref{ua1}) tells us that whether $\chi_\pi - \chi_\delta$ vanishes in the chiral limit or not depends not only on $\rho(0,0)=0$
but also on the approach of $\rho(\lambda,m)$ to the origin in the limits of $m \rightarrow 0$ and $\lambda \rightarrow 0$.
Since only eigenvalues around $\lambda=0$ can contribute to the integral (\ref{ua1}) when $m \rightarrow 0$ is taken, 
one possibility to let $\chi_\pi - \chi_\delta$ become zero is that $\rho(\lambda,m)$ has a gap around $\lambda=0$.

Here, to consider other possibilities, we assume that the dominant contribution to $\rho(\lambda,m)$ at small eigenvalues and quark masses is of the form
$cm^{\alpha-\beta}\lambda^{\beta}$.
In this case the above two quantities are approximately rewritten as
\begin{equation}\label{power_law}
\langle \bar{\psi}\psi \rangle \simeq 2cm^{\alpha}I_1(\beta), \quad \chi_\pi - \chi_\delta \simeq 4cm^{\alpha-1}I_2(\beta),
\end{equation}
with
\begin{equation}
I_k(\beta) \equiv \int^{\Lambda/m}_0 \frac{x^{\beta}}{(1+x^2)^k} dx.
\end{equation}
In this study we examine whether the eigenvalue distribution $\rho(\lambda,m)$ remains finite at $\lambda=0$, i.e. we consider the case $\beta=0$.
The dominant low eigenvalue structure of $\rho(0,m)$ thus is assumed to be $\rho(0,m) = cm^\alpha$.
It is evident from (\ref{power_law}) that we need $\alpha \geq 1$ in order to have $\chi_{\pi}-\chi_{\delta}=\chi_{\mathrm{disc}}$ finite above
the critical temperature. Moreover, only for $\alpha=1$ does $\chi_{\pi}-\chi_{\delta}=\chi_{\mathrm{disc}}$ stay non-zero and thus would provide
a sufficient signature for $U_A(1)$ symmetry breaking. Of course, $\alpha>1$ or $c\equiv 0$ are viable possibilities.
In that case any remnant $U_A(1)$ symmetry breaking would have to arise either from the structure of $\rho(\lambda,m)$ at $\lambda>0$ or a possible
zero mode contribution $\rho(\lambda,m) \sim m^2 \delta(\lambda)$.

\section{Numerical results\label{sec:results}}

\begin{table}[tbp]
\caption{The parameters and statistics of the numerical simulations. $\beta$, $a$, $T$, $m_s$, $m_l$, $N_s$ and $N_\tau$ are the lattice gauge coupling,
lattice spacing, temperature, bare strange quark mass, bare light quark mass, spacial lattice size and temporal lattice size, respectively.
\label{parameters}}
\begin{center}
\begin{tabular}{cccccccc}
\hline \hline
$\beta$ & $a$ [fm]& $ T$ [MeV] & $m_s a$ & \multicolumn{4}{c}{\# configurations} \\ \cline{5-8}
        &           &          &         & $m_l/m_s=1/10$ & 1/20 & 1/27 & 1/40 \\ \hline
\multicolumn{7}{c}{$N^3_s \times N_\tau = 32^3 \times 8$}                      \\ \hline
6.195   & 0.1932    & 127.7    & 0.0880  &  -             & 702  & -    & 360  \\
6.245   & 0.1832    & 134.6    & 0.0830  &  -             & 792  & -    & 822  \\
6.260   & 0.1803    & 136.8    & 0.0810  &  -             & 768  & -    & 834  \\
6.285   & 0.1756    & 140.4    & 0.0790  &  -             & 714  & -    & 882  \\
6.315   & 0.1702    & 144.9    & 0.0760  &  -             & 684  & -    & 936  \\
6.341   & 0.1656    & 148.9    & 0.0740  &  -             & 612  & -    & 402  \\
6.354   & 0.1634    & 151.0    & 0.0728  &  -             & 534  & 1158 & 600  \\
6.390   & 0.1573    & 156.7    & 0.0694  &  -             & 600  & 1458 & 600  \\
6.423   & 0.1520    & 162.2    & 0.0670  & 312            & 696  & 1362 & 540  \\
6.445   & 0.1486    & 165.9    & 0.0652  &  -             & 600  & 1146 & 522  \\
6.460   & 0.1463    & 168.5    & 0.0640  &  -             & 582  & -    &  -   \\
6.488   & 0.1422    & 173.4    & 0.0620  &  -             & 678  & -    &  -   \\
6.550   & 0.1335    & 184.8    & 0.0582  &  -             & 198  & -    &  -   \\
7.150   & 0.0747    & 330.1    & 0.0320  &  -             & 594  & -    &  -   \\ \hline
\multicolumn{7}{c}{$48^3 \times 8$}                                            \\ \hline
6.423   & 0.1520    & 162.2    & 0.0670  &  -             & 144  & -    &  -   \\
6.445   & 0.1486    & 165.9    & 0.0652  &  -             & 156  & -    &  -   \\
7.150   & 0.0747    & 330.1    & 0.0320  &  -             & 306  & -    &  -   \\
\hline \hline
\end{tabular}
\end{center}
\end{table}

We use the tree level improved gauge action and the HISQ action \cite{HISQ} which can reduce cutoff effects due to the taste-symmetry violation
better than any other staggered type quark action.
Part of our gauge configurations have been generated by the HotQCD collaboration \cite{HotQCD}.
We perform our simulations mainly on $32^3 \times 8$ lattices and for a few temperatures on $48^3 \times 8$ lattices to investigate the volume dependence.
The scale was set from the kaon decay constant $f_K$.
Our strange quark mass $m_s$ is fixed to its physical value and we have several light quark masses $m_l$ which have values of
$m_s/40$, $m_s/27$, $m_s/20$ and $m_s/10$ corresponding to the lightest (Goldstone) pion of about 115, 140, 160 and 230 MeV, respectively.
Here $m_l=m_s/27$ gives an almost physical pion mass and the corresponding pseudocritical temperature in the continuum limit
has been estimated at 154(9) MeV by using $O(4)$ scaling fits to the chiral condensate and susceptibility.
The details of the determination of the lattice spacing, the strange quark mass and the pseudocritical temperature have been discussed in Ref.~\cite{HotQCD}.
Each 10th trajectory is chosen for measurements after skipping at least 500 trajectories for thermalization.
Statistical errors are estimated by the jackknife method.
Our simulation parameters and statistics are summarized in Table \ref{parameters}.

We calculated the lowest 100 positive eigenvalues of the staggered (HISQ) Dirac operator for all parameter sets
since the staggered Dirac operator is anti-hermitian and has only pure imaginary eigenvalues in complex conjugate pairs.
We evaluate the eigenvalue density by binning eigenvalues in some small intervals for each configuration
and normalize it so that $\int d \lambda \rho(\lambda,m)=$ $(T/V)\times$\# eigenvalues.

\begin{figure}[tbp]
 \begin{center}
  \includegraphics[width=52.5mm, angle=-90]{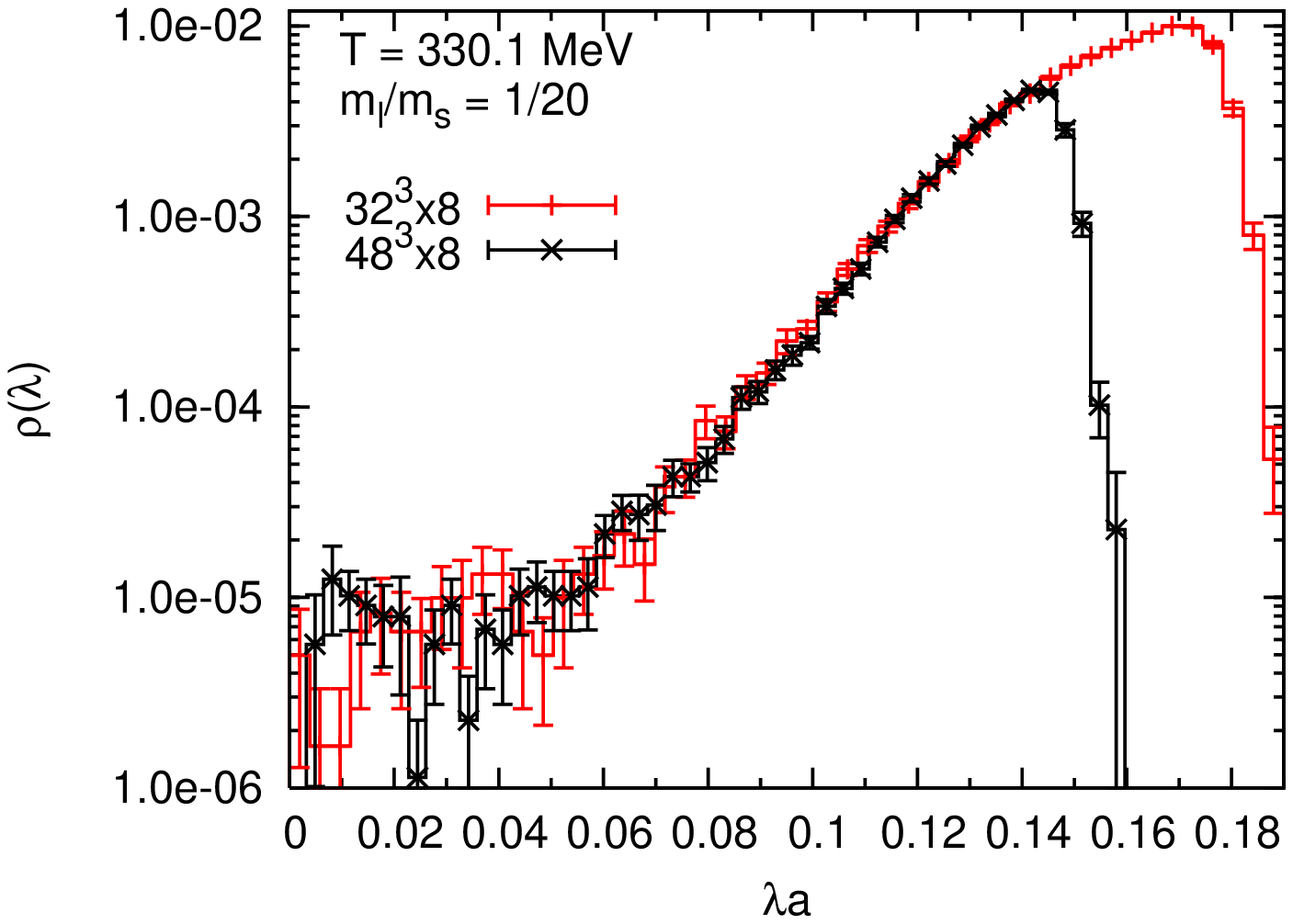}
  \includegraphics[width=52.5mm, angle=-90]{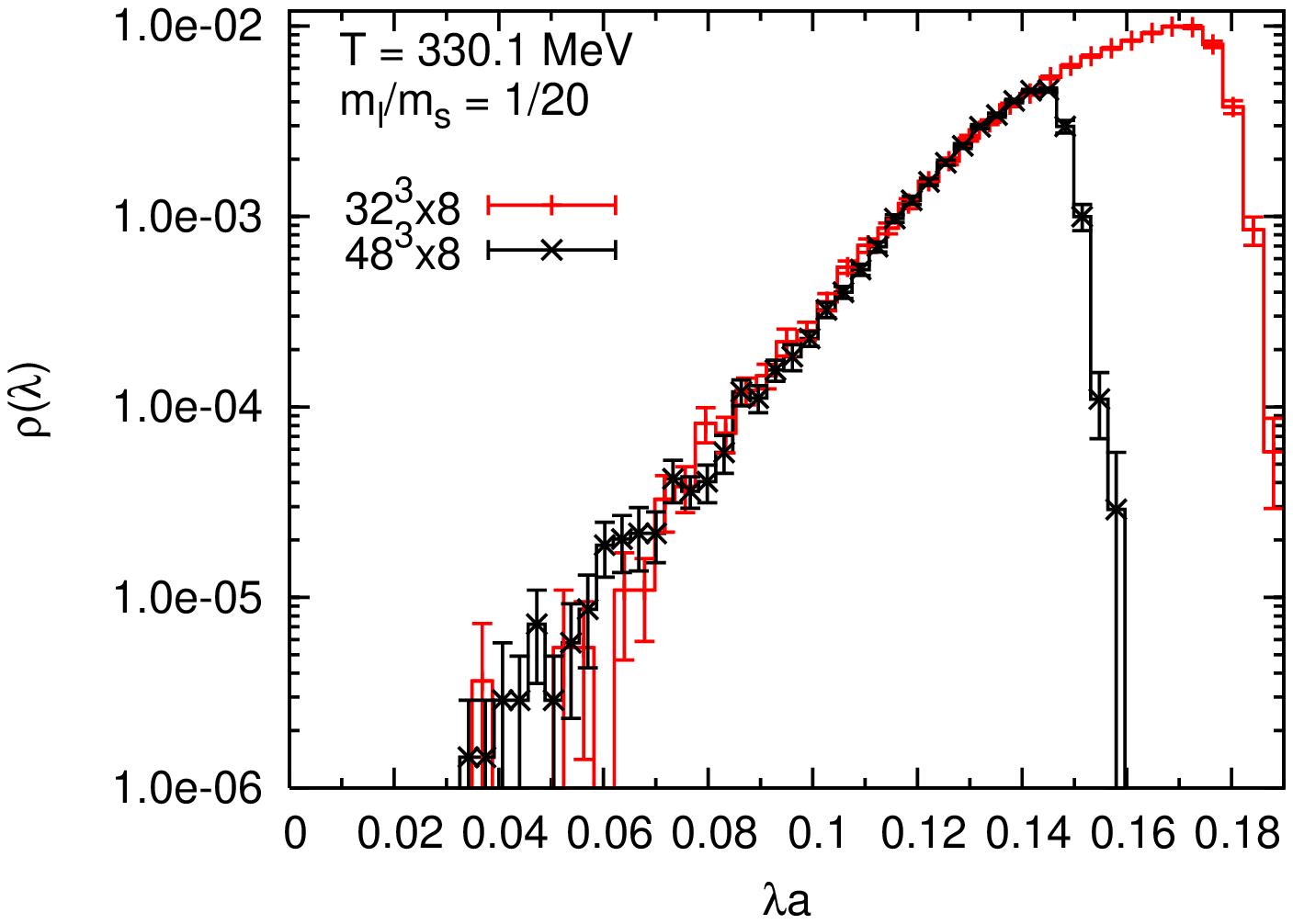}
  \caption{Volume dependence of the Dirac eigenvalue density for $m_l/m_s=1/20$ at $T=330.1$ MeV for all configurations (left) and configurations with $Q=0$ (right). \label{density_Vdep}}
 \end{center}
\end{figure}

In our previous report \cite{lattice2011} we found suppression of $\rho(\lambda,m)$ around $\lambda=0$ and lack of eigenvalues very close to zero above $T_c$.
At the same time, there was also a tail approaching the origin in $\rho(\lambda,m)$ even at our highest temperature, i.e. 330.1 MeV.
Since small eigenvalues could suffer from finite volume as well as topological effects which should vanish in the thermodynamic limit,
investigating the volume dependence of $\rho(\lambda,m)$ is needed to judge whether there is a real gap around $\lambda=0$ in $\rho(\lambda,m)$.
In the left side of Fig.~\ref{density_Vdep} the volume dependence of $\rho(\lambda,m)$ at $T=330.1$ MeV is shown;
$\rho(\lambda,m)$ seems to have a quite small volume dependence in a region of $\lambda<0.05$, although statistics is not enough to see it clearly.
We also check for the effect from the (would-be) zero modes of topological non-trivial configurations whose contribution to the eigenvalue density at zero would vanish in the thermodynamic limit.
The right side of Fig.~\ref{density_Vdep} shows the same result as on the left side but for topological trivial configurations.
Here the topological charge $Q$ is determined form a lattice $F\tilde{F}$ operator with HYP fat links \cite{Topological_charge}.
It is clearly found that contribution form small eigenvalues is more strongly suppressed and eigenvalues less than $\lambda \simeq 0.03$ disappears in this case.
However, since $\rho(\lambda,m)$ seems to decrease continuously and exponentially, it is possible that the tail of $\rho(\lambda,m)$
continues below the lowest value of our $\rho(\lambda,m)$, when a more precise determination is made.
Increased statistics and bigger volumes are needed to make more definite statements.

\begin{figure}[tbp]
 \begin{center}
  \includegraphics[width=52.5mm, angle=-90]{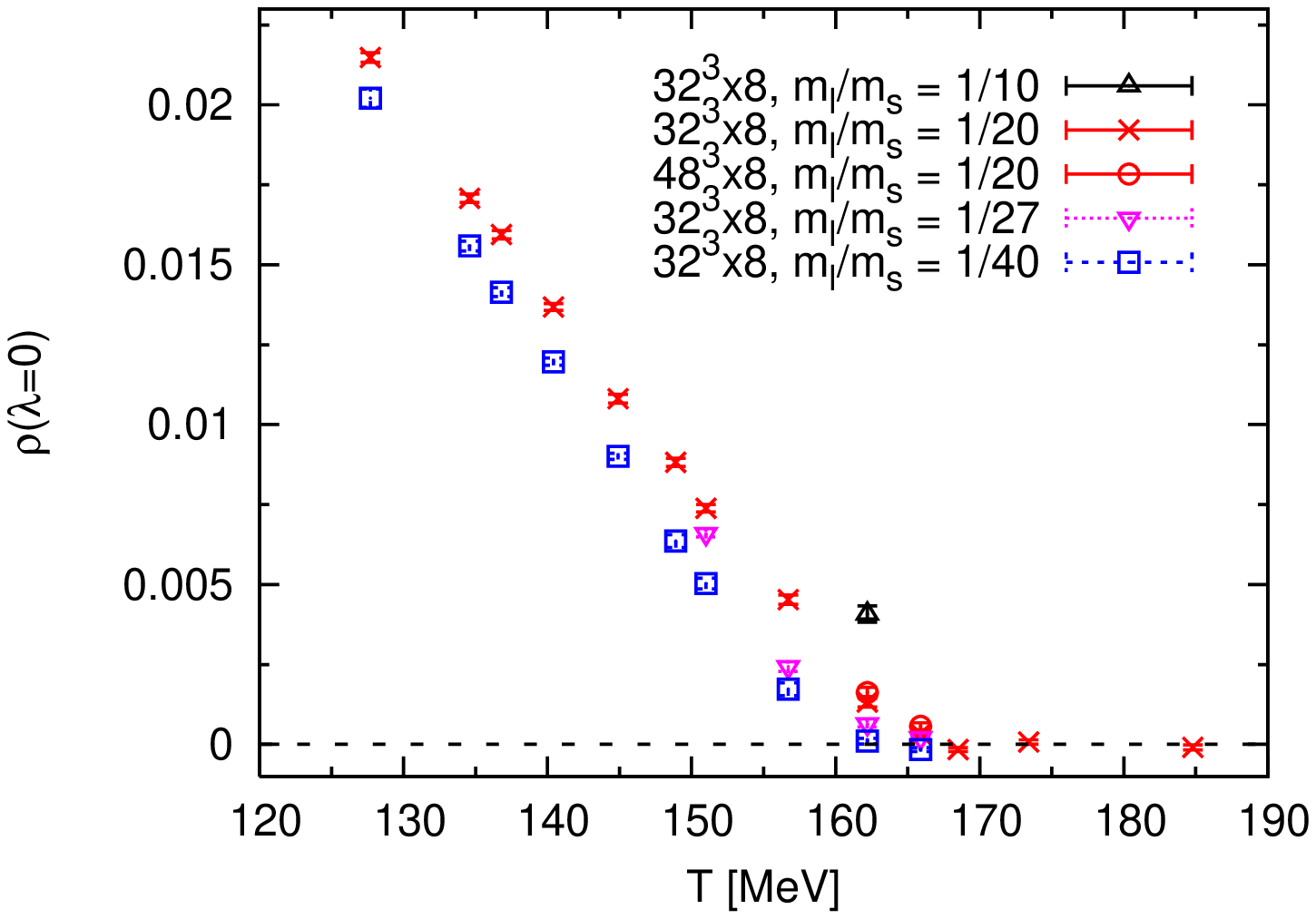}
  \includegraphics[width=52.5mm, angle=-90]{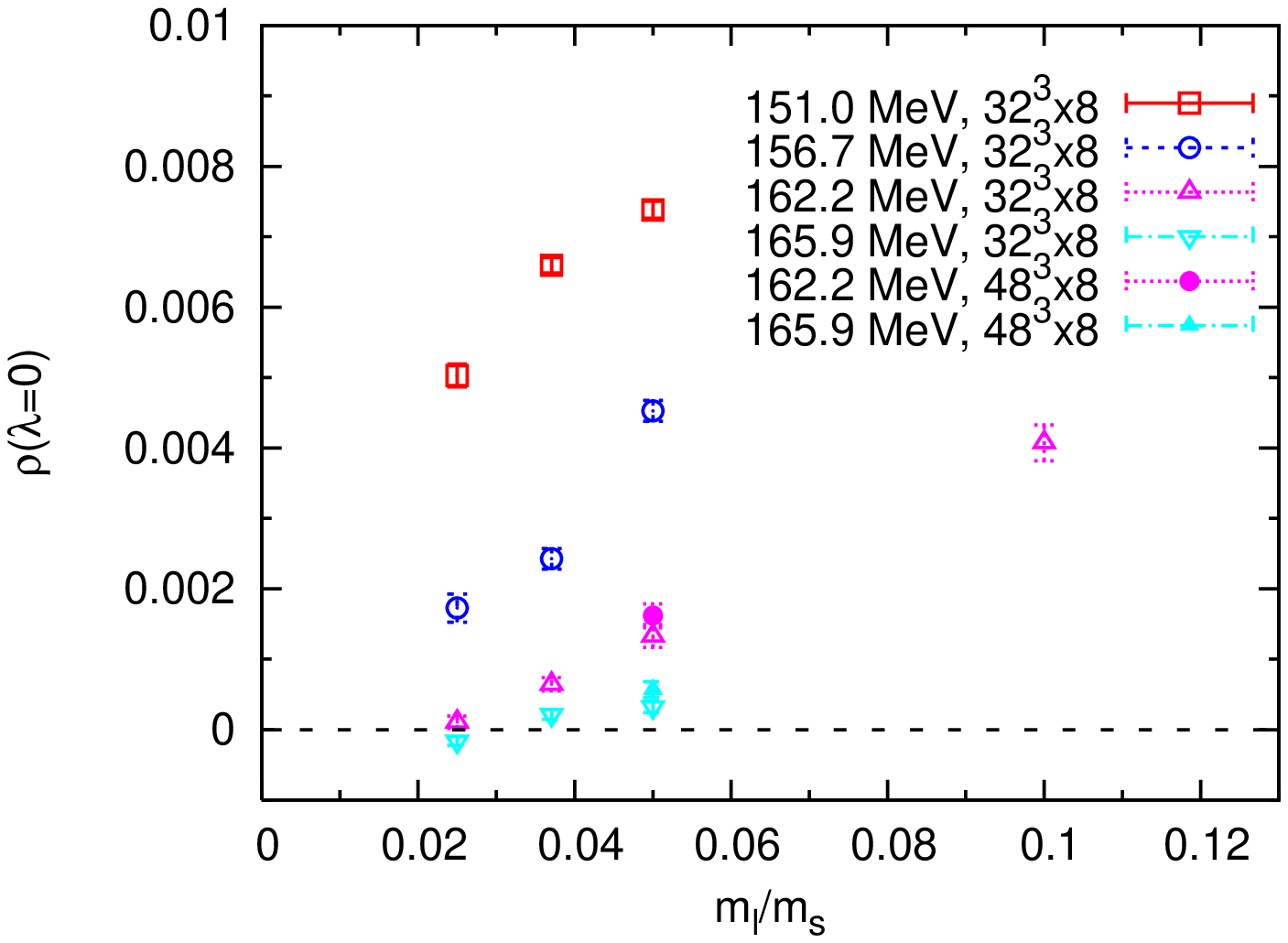}
  \caption{The temperature dependence (left) and the quark mass dependence (right) of the Dirac eigenvalue density at $\lambda=0$.
  	\label{rho0}}
 \end{center}
\end{figure}

As discussed in the previous section, the quark mass dependence of $\rho(0,m)$ is also important to understand $U_A(1)$ restoration above $T_c$.
To estimate $\rho(0,m)$ we fit $\rho(\lambda,m)$ to a cubic polynomial.
In this fit analysis we choose the same bin size of 0.0005 for all eigenvalue densities which we calculated
in order to control a systematic uncertainty coming from the bin size dependence.
The lower bound of the fit range is fixed to the lowest bin and the upper bound is chosen so that $\chi^2$/dof is as close to unity as possible.
Then our $\chi^2$/dof is in the range from 0.88 to 2.1. In the left side of Fig.~\ref{rho0} the temperature dependence of $\rho(0,m)$ is shown.
It can be seen that $\rho(0,m)$ monotonically decreases as the temperature increases and, within statistical errors, it seems to vanish around $T=160$ -- 170 MeV
at least for $m_l/m_s=1/40$, 1/27 and 1/20.
Here we note that even above the crossover transition temperature $T_c$, as determined from the chiral condensate or susceptibility, the condensate will stay
non-zero for non-zero quark masses. The eigenvalue density $\rho(0,m)$ thus may stay non-zero and approaches zero smoothly also above $T_c$.
The right side of Fig.~\ref{rho0} shows the quark mass dependence of $\rho(0,m)$ at four temperatures around $T_c$, namely $T=151.0$, 156.7, 162.2 and 165.9 MeV.
At first, $\rho(0,m)$ seems to approach zero in the chiral limit at the lower two temperatures as expected in the chiral symmetric phase.
However it is difficult to determine whether $\rho(0,m)$ approaches zero linearly so far. To do so we need simulations with smaller quark mass.
On the other hand, data for the higher two temperatures are very close to zero already at $m_l/m_s=1/40$.
However, data for two different volumes at $m_l/m_s=1/20$ show finite volume effects
which indicate an increase of the density in the thermodynamic limit. This means that our results at $T=166.8$ and 170.5 MeV underestimate
the infinite volume result. We thus need larger lattices to get more accurate results.
In addition, the eigenvalue density around zero at temperatures close to $T_c$ is sensitive to fit Ansatze since it changes quickly with respect to $\lambda$.
Hence using any other fit Ansatze will help to estimate systematic uncertainty.

\section{Conclusions\label{sec:conclusions}}

We study $U_A(1)$ symmetry breaking in the chiral symmetric phase in terms of the Dirac eigenvalue distribution.
In our (2+1)-flavor dynamical simulations, using the tree level improved gauge action and the HISQ action, we investigate the volume and
quark mass dependence of the eigenvalue density.

Even in the eigenvalue density at $T=330.1$ MeV, we find that there is a tail approaching the origin with only a small volume dependence.
Moreover the lower end of the tail vanishes when only topological trivial configurations are used.
However it is not clear whether there is a gap or whether a tail exists and would be visible with more statistics.
We also find that the eigenvalue density at zero seems to go to zero in the chiral limit as expected for an unbroken chiral symmetry,
at $T=151.0$ and 156.7 MeV, which are temperatures around $T_c$. We also try to see, if the eigenvalue density at zero has a linear quark-mass dependence,
which would be sufficient to signal breaking of the $U_A(1)$ symmetry in the chiral symmetric phase. However it is difficult to identify it so far.

Increasing statistics and simulations with larger volumes and smaller quark masses are need to make the above points clear.

\acknowledgments{
This work has been supported in part by contracts No. DE-AC02-98CH10886 with the U.S. Department of Energy
and the Bundesministerium f\"{u}r Bildung und Forschung under grant 06BI9001.
The numerical simulations have been performed on the BlueGene/L at the New York Center for Computational
Sciences (NYCCS) which is supported by the U.S. Department of Energy and by the State of New York and
the infiniband cluster of USQCD at Jefferson Laboratory.
This work has been in part based on the MILC collaboration's public lattice gauge theory code \cite{MILC_code}.
}

\end{document}